\documentclass[a4paper,11pt]{article}
\usepackage{jcappub} % for details on the use of the package, please see the JINST-author-manual
\usepackage{bm}

\title{\boldmath Prompt cusps in hierarchical dark matter halos: Implications for annihilation boost}

\author[a,b]{Shin'ichiro Ando,}
\author[a, c]{Mart\'in Mor\'o Gonz{\' a}lez}
\author[a]{and Youyou Li}
\affiliation[a]{GRAPPA Institute, University of Amsterdam, Science Park, 1098 XH Amsterdam, The Netherlands}
\affiliation[b]{Kavli Institute for the Physics and Mathematics of the Universe, University of Tokyo, Chiba 277-8583, Japan}
\affiliation[c]{Centro de Investigaciones Energéticas, Medioambientales y Tecnológicas (CIEMAT), E-28040 Madrid, Spain}

\emailAdd{s.ando@uva.nl, martin.moro@ciemat.es, y.li4@uva.nl}

\abstract{Recent simulations have identified long-lived ``prompt cusps''---compact remnants of early
density peaks with inner profiles $\rho\propto r^{-3/2}$.
They can survive hierarchical
assembly and potentially enhance signals of dark matter annihilation.
In this work, we incorporate prompt cusps into the semi-analytic substructure framework
\textsc{SASHIMI}, enabling a fully hierarchical, environment-dependent calculation of the
annihilation luminosity that consistently tracks subhalos, sub-subhalos, and tidal stripping.
We assign prompt cusps to first-generation microhalos and propagate their survival through the
merger history, including an explicit treatment of cusps associated with stripped substructure.
We find that the substructure hierarchy converges rapidly once a few levels are included, and
that prompt cusps can raise the total annihilation boost of Milky-Way--size hosts at $z=0$ to
$B\sim 50$ for fiducial cusp-occupation assumptions, compared to a subhalo-only
baseline of $B_{\rm sh}\sim\mathrm{few}$.
Across a wide range of host masses and redshifts, prompt cusps increase the normalization of
$B(M_{\rm host},z)$ while largely preserving its mass and redshift trends.
Compared to universal-average, peak-based estimates, our fiducial boosts are lower by about a factor of a few, primarily reflecting a correspondingly smaller inferred cusp abundance in
host halos, highlighting the importance of unifying peak-based cusp formation with
merger-tree evolution and environmental dependence.}

\begin{document}
\maketitle
\flushbottom

\section{Introduction}
\label{sec:Introduction}

Cosmological observations across a wide range of scales indicate that more than 80\% of the matter content of the Universe is in the form of non-luminous, non-baryonic dark matter.
Despite this strong empirical support (from the cosmic microwave background, large-scale structure, and gravitational lensing), the microscopic nature of dark matter remains unknown.
In the standard cold dark matter (CDM) paradigm, dark matter is assumed to be non-relativistic during structure formation, leading naturally to hierarchical clustering: the smallest bound objects collapse first and subsequently merge to form larger halos.
If dark matter consists of weakly interacting massive particles (WIMPs), such as the supersymmetric neutralino, the earliest gravitationally bound halos can be extremely small, with masses as low as ${M} \sim 10^{-6}\,{M_\odot}$ and extending up to cluster scales of ${M} \sim 10^{15}\,{M_\odot}$~\cite{Hofmann_2001, PhysRevLett.97.031301, Bertschinger:2006nq, Green:2003un, Loeb:2005pm,Diamanti:2015kma}.
As these small halos form at high redshift and merge into larger hosts, many are tidally disrupted while others survive as long-lived bound remnants---commonly referred to as subhalos---embedded within the larger structures that comprise the cosmic web.

The abundance of substructure within CDM halos has important observational implications.
Self-bound subhalos significantly enhance the annihilation rate of WIMP dark matter, boosting the resulting gamma-ray luminosity compared to that of a smooth host halo. This enhancement has motivated extensive studies on the detectability of annihilation signals and the role of substructure in shaping the diffuse gamma-ray background~\cite{Silk:1992bh, Bergstrom:1998jj, Bergstrom:1998zs, Ando:2005hr, Pieri:2007ir, Koushiappas:2003bn, Stoehr:2003hf, Berezinsky:2006qm, Lavalle:2007apj, Bartels:2015uba, Hiroshima:2018kfv, Ando_2019, Ishiyama:2019hmh}.
A key ingredient in such predictions is the assumed inner density profile of halos and subhalos.

Numerical simulations have long indicated that CDM halos follow a nearly universal density profile, commonly parametrized by the Navarro-Frenk-White (NFW) form, $\rho(r)\propto r^{-1}$ in the inner regions and $\rho(r)\propto r^{-3}$ at large radii~\cite{Navarro_1997}.
However, recent high-resolution simulations have revealed a qualitatively different behavior in the earliest-forming halos.
Small-scale peaks in the initial density field can undergo rapid, nearly self-similar collapse, producing extremely steep inner cusps with $\rho(r)\propto r^{-1.5}$---significantly steeper than the NFW profile~\cite{Ishiyama:2010es, Anderhalden:2013wd, Ishiyama:2014uoa, Polisensky:2015eya,Ogiya:2017hbr,White:2022yoc, 10.1093/mnras/stac3373,Ondaro-Mallea:2023qat, delos2024limits, delos2023prompt, Ginat:2025kuz}.
These prompt cusps appear generically in the first generation of halos, with their number and properties determined by the statistics of primordial density peaks.
Only peaks with sufficiently low ellipticity and without strong interactions with neighboring structures form such cusps; others relax to an NFW-like structure instead.
Importantly, simulations suggest that once formed, prompt cusps tend to survive subsequent mergers and accretion events, remaining as long-lived features embedded in the inner regions of larger halos~\cite{Ogiya:2016hyo, Angulo:2016qof, 10.1093/mnras/stac3373, Wang:2025wki}.

The existence of prompt cusps has significant implications for indirect searches for dark matter. 
Because the annihilation rate scales with the square of the local dark matter density, the extremely steep 
$r^{-1.5}$ profiles generated by the rapid collapse of small-scale peaks 
can substantially enhance the gamma-ray emissivity associated with WIMP annihilation. 
Delos and White~\cite{delos2023prompt} demonstrated that, for thermally produced WIMPs, every solar mass of dark matter 
may contain thousands of Earth-mass prompt cusps, leading to an overall annihilation rate that exceeds 
predictions based on smooth-halo or traditional subhalo models by at least an order of magnitude. 
Their analysis further showed that prompt-cusp emission is less centrally concentrated than the conventional 
NFW-based expectation, thereby modifying the relative importance of diffuse components such as the 
extragalactic gamma-ray background~\cite{delos2023prompt}. 
Subsequent works test these predictions against observations. 
Constraints from the isotropic gamma-ray background measured by the \textit{Fermi}-LAT have been shown 
to strongly limit the viable parameter space for WIMP annihilation once prompt cusps are included~\cite{delos2024limits}. 
Additional studies have considered environmental effects such as disruption by stars and tidal fields, 
finding that cusp survival can significantly influence the expected annihilation signal~\cite{stucker2023effect}. 
Observations of galaxy clusters have been used to place further constraints on 
prompt-cusp annihilation scenarios, in some cases suggesting substantial tension between 
cusp-enhanced predictions and current gamma-ray limits~\cite{Crnogorcevic:2025nwp, Olea-Romacho:2025qag}. 

Importantly, nearly all existing analyses characterize the contribution of prompt cusps using a 
\textit{universal average} boost factor; they assume that cusp properties and survival 
statistics are identical across cosmic environments. 
While this approach provides a valuable first estimate, it cannot capture the potential dependence 
of annihilation enhancement on host-halo mass, redshift, merger history, or the hierarchical 
structure of subhalos and their internal evolution. 
To address these shortcomings, in this work we incorporate prompt-cusp physics directly into the 
semi-analytic substructure framework \textsc{SASHIMI}~\cite{Hiroshima:2018kfv},\footnote{\url{https://github.com/shinichiroando/sashimi-c}} enabling a fully hierarchical and 
environment-dependent treatment of annihilation signals.
(For \textsc{SASHIMI} for other non-cold dark matter candidates, see, e.g., Refs.~\cite{Dekker:2021scf, Ando:2024kpk}.)
Specifically, we embed one prompt cusp into every first-generation subhalo generated by 
\textsc{SASHIMI}, and we track its survival within the full hierarchy of substructures, including 
sub-subhalos and even smaller progenitors. 
Cusp disruption is modeled through a parameterization that captures the tidal evolution of each 
host–subhalo system, allowing us to propagate survival probabilities consistently across the 
merger history. 
With this enhanced model, we re-evaluate the annihilation boost factor for a wide range of host-halo 
masses and find that Milky-Way–like halos generically exhibit total boosts of order tens---comparable 
to those inferred in previous universal-average studies~\cite{delos2023prompt}, but now obtained through a physically 
grounded, self-consistent hierarchical calculation. 
We also release an updated version of the public \textsc{SASHIMI} package, 
which includes our prompt-cusp implementation and associated tools for annihilation boost calculations.

The remainder of this paper is organized as follows. 
In Sec.~\ref{sec:prompt cusps}, we summarize the formation of prompt cusps and the physical 
quantities that characterize them. 
Section~\ref{sec:halos} reviews the \textsc{SASHIMI} framework and describes how we model the 
hierarchical population of subhalos and sub-subhalos, by embedding prompt cusps. 
In Secs.~\ref{sec:Results}, we present the evolution and survival of prompt 
cusps within host halos and incorporate their contribution into the annihilation signal by embedding 
one cusp into each first-generation subhalo. 
We then evaluate the resulting annihilation boost across host-halo masses and compare our findings 
with previous studies. 
Finally, Sec.~\ref{sec:Discussion} summarizes our conclusions and discusses prospects for future work.

\section{Prompt cusps}
\label{sec:prompt cusps}

The theoretical framework adopted in this work follows the analysis of Refs.~\cite{PhysRevD.100.023523, 10.1093/mnras/stac3373, delos2023prompt}, which showed that small-scale peaks in the primordial 
density field undergo rapid, nearly self-similar collapse and form extremely steep density cusps. 
These structures arise at the earliest epochs of hierarchical structure formation, prior to 
virialization into the familiar NFW profile, and retain a characteristic inner 
slope $\rho \propto r^{-1.5}$ that persists unless disrupted by subsequent mergers or encounters.  
In this section, we summarize the essential physics governing the formation of prompt cusps, their 
characteristic scales, and the resulting annihilation luminosity.

\subsection{Formation of cusps from primordial density peaks}

Density fluctuations in the early universe are described by the matter power spectrum $P(k)$, which can be written as $P(k) \propto k^{n_s}\,T^{2}(k)$, where $n_s \approx 0.965$ is the power-law index of the primordial curvature perturbation~\cite{refId0} and $T(k)$ is the transfer function encoding the suppression of small-scale modes and the growth of perturbations prior to matter–radiation equality~\cite{Bardeen:1985tr, Eisenstein:1997ik}.  
The dimensionless form of the spectrum is $\mathcal{P}(k) = {k^{3}} P(k)/{(2\pi^{2})}$,
and its time evolution is governed by the linear growth factor $D(z)$: $\mathcal{P}(k,z) = D^{2}(z)\,\mathcal{P}(k)|_{z=0}$.
For a thermal WIMP with mass $m_\chi = 100~\mathrm{GeV}$ and kinetic decoupling temperature 
$T_{\mathrm{kd}} = 30~\mathrm{MeV}$, small-scale power is suppressed below the free-streaming scale of $k_{\rm fs} = 1.06\times 10^{6} \, \mathrm{Mpc}^{-1}$~\cite{Green:2003un}, by multiplying $\mathcal P(k,z)$ further by $\exp(-k^2/k_{\rm fs}^2)$.

Peaks in the primordial density field, $\delta(\bm{x}) = (\rho - \bar\rho)/\bar\rho$, collapse first 
along their steepest directions.  
For sufficiently low ellipticity, the collapse proceeds in a nearly radial fashion and produces a 
self-similar density profile,
\begin{equation}
    \rho(r) = A\,r^{-3/2},
    \label{eq:cusp profile}
\end{equation}
where the amplitude $A$ depends on the peak height and curvature.  
Delos and White~\cite{10.1093/mnras/stac3373, delos2023prompt} showed that
\begin{equation}
\label{eq:A_cusp}
    A \simeq 24\,\bar{\rho}_{0}\,a_{\mathrm{coll}}^{-3/2}\,\textcolor{black}{\mathcal R}^{3/2},
\end{equation}
with $\bar\rho_0$ the present-day mean dark matter density and $\textcolor{black}{\mathcal R}$ the characteristic comoving 
curvature scale, $\textcolor{black}{\mathcal R} = |\delta/{\nabla^{2}\delta}|^{1/2}$, where $\delta$ and $\nabla^2\delta$ are the linearly extrapolated density contrast
and curvature at the peak position.
\textcolor{black}{Here, $a_{\rm coll}$ denotes the collapse scale factor of the peak; its evaluation is discussed in the next subsection.}

To characterize peaks in the Gaussian density field with $\delta$ and $\nabla^2\delta$, we define the spectral moments as
\begin{equation}
    \sigma_j^2
      = \int_0^\infty \frac{{\rm d}k}{k}
        k^{2j}\,\mathcal{P}(k) ,
    \qquad j = 0,1,2,
\end{equation}
which fully determine the statistics of peaks.
Following Ref.~\cite{Bardeen:1985tr}, the joint distribution of peak height and 
curvature is parametrized by $\nu \equiv {\delta}/{\sigma_0}$ and $x \equiv -{\nabla^2\delta}/{\sigma_2}$.
We draw $\nu$ and $x$ from the peak 
distribution (see, e.g., Appendix~A of Ref.~\cite{delos2023prompt}), reconstructing the physical peak quantities via $\delta = \nu \sigma_0$, 
$\nabla^2\delta = -x \sigma_2$, to calculate the characteristic comoving scale of the peak $\textcolor{black}{\mathcal R} = |{\delta}/{\nabla^2\delta}|^{1/2}$.

\subsection{Collapse time and characteristic radii}

The collapse scale factor for a peak of height $\delta(a)$ at scale factor $a$ is~\cite{delos2023prompt}
\begin{equation}
\label{eq:a_coll}
    a_{\mathrm{coll}} = 
    \left[ f_{\mathrm{ec}}(e,p)\,\frac{\delta_{\mathrm{c}}}{\delta(a)}\right]^{1/g} a,
\end{equation}
where $f_{\mathrm{ec}}(e,p)$ is the ellipsoidal-collapse correction sampled from a distribution given the peak height $\delta$~\cite{Sheth:1999su}, $\delta_{\mathrm{c}} = 1.686$ is the 
critical threshold for spherical collapse, and $g \simeq 0.901$ is the small-scale growth index.
In order for a cusp to form in each peak, the ellipticity $e$ and prolateness $p$ need to satisfy the condition: $e^2-p|p|<0.26$.
A fraction $f_{\rm coll}$ of all the peaks will collapse to form a cusps in them.

Once formed, the cusp extends out to a physical radius
$r_{\mathrm{cusp}} \simeq 0.11\,a_{\mathrm{coll}}\, \textcolor{black}{\mathcal R}$.
The profile cannot continue to arbitrarily small radii:
the coarse-grained phase-space density of collisionless
dark matter can never exceed its primordial maximum
\cite{Tremaine:1979we}. For a thermal relic, this maximum is
\begin{equation}
\label{eq:fmax}
    f_{\mathrm{max}} =
    \frac{1}{(2\pi)^{3/2}}
    \left( \frac{m_\chi}{T_{\mathrm{kd}}} \right)^{3/2}
    \bar{\rho}_{0}\, a_{\mathrm{kd}}^{-3}.
\end{equation}
The self-similar $\rho(r)=A\,r^{-3/2}$ profile must
therefore flatten at a core radius $r_{\rm core}$, below
which the implied phase-space density would otherwise
exceed $f_{\max}$.  It is useful to introduce the
corresponding core density,
\begin{equation}
    \rho_{\rm core} \equiv \rho(r_{\rm core})
                    = A\, r_{\rm core}^{-3/2},
\end{equation}
which marks the innermost scale at which the cusp is
regulated by the primordial phase-space bound.
The value of \(r_{\rm core}\) is obtained by equating $f_{\rm max}$ with the 
phase-space density implied by the power-law cusp, yielding
\begin{equation}
\label{eq:r_core_definition}
    r_{\rm core} \simeq 
    \left[
        3\times10^{-5} G^{-3} f_{\max}^{-2} A^{-1}
    \right]^{2/9}.
\end{equation}
This core radius plays a central role in determining the annihilation luminosity of a prompt cusp, 
since the integral of \(\rho^2\) is dominated by the region between \(r_{\rm core}\) and 
\(r_{\rm cusp}\).  

\subsection{Annihilation luminosity of a prompt cusp}

Because the annihilation rate scales as $\rho^{2}$, prompt cusps are exceptionally bright sources 
compared to NFW-like inner profiles.  
Integrating the density-squared profile gives the 
annihilation ``luminosity parameter''~\cite{stucker2023effect}
\begin{equation}
\label{eq:Jcusp}
    J_{\mathrm{cusp}} 
    = \int 4\pi r^{2}\rho^{2}(r)\,dr 
    = 4\pi A^{2}
    \left[ 0.531 + \ln\!\left(\frac{r_{\mathrm{cusp}}}{r_{\mathrm{core}}}\right) \right].
\end{equation}
The quantity $J_{\mathrm{cusp}}$ appears repeatedly throughout this work and forms the basic unit of 
annihilation luminosity for each prompt cusp before any tidal evolution or survival probability is 
applied.

\subsection{Prompt cusp properties and annihilation enhancement}
\label{subsec:promptcusps}

The properties of individual prompt cusps are determined by a small set of peak parameters,
namely the peak height $\nu$, curvature $x$, and the ellipticity--prolateness parameters $(e,p)$.
These quantities are sampled from their joint probability distributions as prescribed by
peak statistics~\cite{Bardeen:1985tr}.
All remaining cusp properties then follow deterministically from $(\nu,x,e,p)$ through the
collapse condition and subsequent dynamical relations.

Using this sampling procedure, we compute ensemble-averaged properties of prompt dark matter
cusps. Averaging over the population of collapsing peaks, we obtain the following representative
values:
\begin{align}
\langle \textcolor{black}{\mathcal R}\rangle &\simeq 1.5~\mathrm{pc}, \\
\langle a_{\rm coll}\rangle &\simeq 0.09, \\
\langle A\rangle &\simeq 1.5\times 10^{-4}\,M_\odot\,\mathrm{pc^{-1.5}}, \\
\langle r_{\rm core}\rangle &\simeq 1.9\times 10^{-5}~\mathrm{pc}, \\
\langle r_{\rm cusp}\rangle &\simeq 1.2\times 10^{-2}~\mathrm{pc}, \\
\langle M_{\rm cusp}\rangle &\simeq 9\times 10^{-7}\,M_\odot ,\\
f_{\rm coll} &\simeq 0.48,
\end{align}
which are in good agreement with those obtained by Ref.~\cite{delos2023prompt}.

The characteristic curvature scale $\textcolor{black}{\mathcal R}$ and the collapse scale factor $a_{\rm coll}$ indicate that prompt cusps form at very early epochs, well before nonlinear structure formation on galactic scales.
Roughly half of the peaks that satisfy the collapse condition proceed to form cusps, with a fraction $f_{\rm coll}\simeq 0.48$.
The resulting cusps are highly compact objects with a steep inner density profile, characterized by a small core radius $r_{\rm core}\ll r_{\rm cusp}$.  
The normalization parameter $A$ sets the amplitude of the density profile [Eq.~(\ref{eq:cusp profile})],
which leads to a total cusp mass well below stellar scales, $M_{\rm cusp}\sim 10^{-6}M_\odot$.
Despite their small mass, the extremely high central densities result in a significant enhancement of the dark matter annihilation signal.  
The mean annihilation factor per cusp is found to be
\begin{equation}
\langle J_{\rm cusp}\rangle \simeq 4.7\times 10^{-6}\,M_\odot^2\,\mathrm{pc^{-3}},
\label{eq:average Jcusp}
\end{equation}
demonstrating that prompt cusps can provide a non-negligible contribution to annihilation observables once their abundance is taken into account.

\section{Semi-analytical treatment of halo substructure and prompt cusps}
\label{sec:halos}

\subsection{SASHIMI: weighted subhalo catalog}

To model the contribution of substructure to dark matter annihilation signals,
we employ the Semi-Analytical SubHalo Inference ModelIng (SASHIMI; \cite{Hiroshima:2018kfv, Ando_2019}).
SASHIMI provides an analytic realization of a subhalo population within a host halo
by sampling the subhalo mass-accretion history and its subsequent tidal evolution.
Rather than generating individual halos in an $N$-body sense, the method constructs
a weighted catalog of subhalos whose ensemble averages reproduce the underlying
distribution functions.
In this work, we compute the rms overdensity $\sigma(M)$ from a linear matter power spectrum
that includes a small-scale free-streaming suppression,
$P(k)\rightarrow P(k)\exp(-k^2/k_{\rm fs}^2)$.
Following common practice for suppressed-small-scale-power scenarios, we evaluate
$\sigma(M)$ using a $k$-space top-hat (sharp-$k$) filter, with an upper cutoff
$k_{\max}=\alpha/\textcolor{black}{R_M}$ (where \textcolor{black}{$R_M$} is the Lagrangian radius associated with mass $M$),
which is known to provide a robust mapping between the cutoff scale and the halo mass function
in excursion-set calculations \cite{Schneider:2013ria}. 
The parameter $\alpha$ encodes the residual ambiguity of this mapping; we fix it to
$\alpha=1.8$ by matching $\sigma(M)$ to the real-space top-hat result in the large-mass regime %\MM{Is $\sigma(M)$ calculated within SASHIMI?}.
The resulting $\sigma(M)$ is then used consistently in the extended-Press-Schechter (EPS)-based ingredients of SASHIMI.

Each subhalo in the catalog, labeled by an index $i$, is assigned a set of structural
parameters $(\rho_{s,i}, r_{s,i}, r_{t,i})$, where $\rho_{s,i}$ and $r_{s,i}$ are the
characteristic density and scale radius of an NFW profile, and $r_{t,i}$ is the
truncation radius induced by tidal stripping.
In addition, each entry carries a weight factor $w_i$, which represents the number
of physical subhalos corresponding to that realization.
As a result, integrals over the distribution of subhalo properties,
can be evaluated as weighted sums over the catalog,
\begin{equation}
    \int d\bm{\theta}\,\frac{dN_{\rm sh}}{d\bm\theta}\,(\cdots)
    \;\longrightarrow\;
    \sum_i w_i\,(\cdots)_i .
\end{equation}
where $\bm\theta = (\rho_s, r_s, r_t, \dots)$ represents the subhalos' density profile parameters among others, and $dN_{\rm sh}/\bm\theta$ is their joint distribution function.
\textcolor{black}{Here, the structural parameters assigned to each catalog entry are evolved after accretion
within the SASHIMI framework.
For the tidal stripping calculation, the subhalo mass-loss equation depends explicitly on
the host-halo mass $M_{\rm host}(z)$ and redshift $z$~\cite{Hiroshima:2018kfv, Ando_2019}, so that the time dependence of the
host environment relevant for tidal mass loss is taken into account.}

For example, the annihilation luminosity parameter of an individual subhalo $i$ is given by
\begin{equation}
    J_{\mathrm{sh},i}
    = \int_0^{r_{t,i}} dr\, 4\pi r^2
      \rho_{\mathrm{sh}}^2(r |\rho_{s,i}, r_{s,i}, r_{t,i}) %\nonumber\\
    = \frac{4\pi}{3}\rho_{s,i}^2 r_{s,i}^3
      \left[1 - \frac{1}{\left(1 + r_{t,i}/r_{s,i}\right)^3}\right],
\end{equation}
where $\rho_{\mathrm{sh}}(r)$ denotes the truncated NFW profile.
In the baseline SASHIMI implementation, the above expression describes the
annihilation luminosity of a single (resolved) subhalo realization.
Additional enhancement from unresolved substructure \emph{within} subhalos
(i.e., sub-subhalos and higher-order levels) can be incorporated in a recursive
manner; we refer the reader to Refs.~\cite{Hiroshima:2018kfv, Ando_2019} for a
detailed treatment and its impact on the boost factor.
In this work, we will revisit the hierarchical contribution explicitly in the
next subsection, where we quantify the effective subhalo counts when including
$\mathrm{sub}^n$-subhalos.
The total contribution from all subhalos in a host halo can then be written as
\begin{equation}
    J_{\mathrm{sh,\,total}}
    = \int d\rho_s \int dr_s\int dr_t\,
      \frac{d^3N_{\rm sh}}{d\rho_s dr_s dr_t}\,J_{\mathrm{sh}}(\rho_s,r_s,r_t)
    \simeq \sum_i w_i\,J_{\mathrm{sh},i}.
    \label{eq:Jsh}
\end{equation}

Using this framework, Refs.~\cite{Hiroshima:2018kfv, Ando_2019}
demonstrated that the annihilation boost factor due to subhalos in a Milky-Way–size
halo is modest, typically $B_{\mathrm{sh}}\sim 2$--$3$.
In the following, we extend this formalism by embedding prompt cusps within each
subhalo realization and reassessing the resulting annihilation enhancement.

\subsection{Hierarchical substructure and effective prompt-cusp counts}
\label{sec:substructure_hierarchy}

As a concrete example, we consider a Milky-Way--size host halo with
$M=10^{12}\,M_\odot$ at $z=0$.
We include the hierarchy of substructure up to $\mathrm{sub}^n$-subhalos.

At zeroth order ($n=0$), only subhalos are present and each subhalo's density profile is assumed to be smooth.
The expected number of subhalos can be written as the sum of weights in the SASHIMI catalog,
\begin{equation}
N_{\rm sh}^{(0)}(M,z) = \sum_{i} w_{i}(M,z),
\end{equation}
which evaluates to $2.7\times 10^{15}$.
Next, at first order ($n=1$), we include sub-subhalos within each host-level subhalo.
The effective number of subhalos is then given by
\begin{equation}
N_{\rm sh}^{(1)}(M,z)
= \sum_{i} w_{i}(M,z)
\left[
1+\sum_{j} w_{j}\left(m_{a,i},z_{a,i}\right)
\right],
\end{equation}
where $m_{a,i}$ and $z_{a,i}$ are the accretion mass and accretion redshift of the
$i$th subhalo, and the inner sum runs over the sub-subhalo catalog conditioned on
$\left(m_{a,i},z_{a,i}\right)$.

Proceeding in the same manner, including sub$^n$-subhalos corresponds to iterating this
construction. For example, up to $n=3$, we write
\begin{eqnarray}
N_{\rm sh}^{(3)}(M,z)
&=& \sum_{i} w_{i}(M,z)
\left\{
1+\sum_{j} w_{j}\left(m_{a,i},z_{a,i}\right)
%\right.\nonumber\\ 
%&&{}\times
\left[
1+\sum_{k} w_{k}\left(m_{a,j},z_{a,j}\right)
\right.\right.\nonumber\\ 
&&\times
\left.\left.
\left(
1+\sum_{l} w_{l}\left(m_{a,k},z_{a,k}\right)
\right)
\right]
\right\}.
\end{eqnarray}
We compute these effective counts up to $n = 4$ for $M=10^{12}M_\odot$ at $z=0$ and summarize the results in Table~\ref{tab:Nsh_hierarchy}.

\begin{table}[t]
\centering
\caption{Effective number of subhalos in a Milky-Way--size host
($M=10^{12}M_\odot$, $z=0$) when including hierarchical substructure up to
$\mathrm{sub}^n$-subhalos.
We additionally report the split into subhalos inside the truncation radius of
their immediate parent, $N_{\rm sh,in}^{(n)}$, and those in the stripped region,
$N_{\rm sh,stripped}^{(n)}$.}
\label{tab:Nsh_hierarchy}
\begin{tabular}{c r r r}
\hline
$n$ &
$N_{\rm sh}^{(n)}$ &
$N_{\rm sh,in}^{(n)}$ &
$N_{\rm sh,stripped}^{(n)}$ \\
\hline
0 & $2.7\times10^{15}$  & $2.7\times10^{15}$ & $0$ \\
1 & $6.6\times10^{15}$  & $3.3\times10^{15}$ & $3.3\times10^{15}$ \\
2 & $9.2\times10^{15}$  & $3.3\times10^{15}$ & $5.9\times10^{15}$ \\
3 & $1.03\times10^{16}$ & $3.3\times10^{15}$ & $7.0\times10^{15}$ \\
4 & $1.06\times10^{16}$ & $3.3\times10^{15}$ & $7.3\times10^{15}$ \\
\hline
\end{tabular}
\end{table}

%In hierarchical structure formation, dark matter (sub)halos can be associated,
%to a good approximation, with peaks of the underlying (filtered) Gaussian density
%field, so that each subhalo realization may be linked to a corresponding peak
%in the initial conditions~\cite{Bardeen:1985tr, Bond:1993we, Bond:1993wd}.
%Motivated by this peak-based picture, we assume that a prompt cusp can form
%around a peak only if the peak collapses sufficiently early, and that only a
%fraction $f_{\rm coll}$ of peaks satisfies this collapse condition (as discussed
%in the previous subsection).  Moreover, not all newly formed prompt cusps are
%expected to survive subsequent evolution: tidal processing and mergers can erase
%a non-negligible fraction of cusps, leading to an overall survival probability
%of $f_{\rm surv}\simeq 0.4$--$0.6$, as suggested by
%cosmological considerations of cusp survival \cite{delos2023prompt}.

In hierarchical structure formation, dark matter halos can be associated,
to a good approximation, with peaks of the underlying Gaussian density
field~\cite{Bardeen:1985tr, Bond:1993we, Bond:1993wd}.
However, not every peak in the initial density field gives rise to a bound halo.
Only peaks that collapse sufficiently early and reach high central densities
form self-bound halos, and it is these collapsed peaks that can host a central
prompt cusp.
Cosmological arguments and numerical studies indicate that roughly half of the
initial peaks satisfy this collapse condition, corresponding to a collapse
fraction $f_{\rm coll}\simeq 0.48$ (see Sec.~\ref{subsec:promptcusps}), while the
remainder never form bound halos and therefore do not contribute to the subhalo
population.
%Once formed, halos subsequently evolve through hierarchical mergers and tidal
%interactions.
%At late times, each bound halo hosts a single central prompt cusp.
In this work, we therefore adopt the physically motivated assumption that every
collapsed primordial halo contains a central prompt cusp.

%Combining these ingredients, we parametrize the fraction of subhalos that host
%a \emph{surviving} prompt cusp as
%\begin{equation}
%    f_{\rm cusp} \equiv f_{\rm coll}\,f_{\rm surv},
%\end{equation}
%such that $f_{\rm cusp}$ may be directly compared to the effective cusp fraction
%adopted in Ref.~\cite{delos2023prompt}.  For reference, our peak-collapse estimate
%gives $f_{\rm coll}\simeq 0.48$, and together with $f_{\rm surv}\simeq 0.4$--$0.6$
%this suggests $f_{\rm cusp}\sim 0.2$--$0.3$, making $f_{\rm cusp}=0.25$ a
%plausible benchmark value.  In what follows, we therefore explore
%$f_{\rm cusp}=\{1,\ 0.5,\ 0.25\}$ as representative choices spanning optimistic
%to conservative scenarios for cusp survival within the subhalo population.

We introduce $f_{\rm cusp}$ as a dimensionless parameter that denotes the
fraction of halos whose central prompt cusps contribute effectively to the
annihilation signal.
Our fiducial choice is $f_{\rm cusp}=1$, corresponding to the assumption that
all collapsed halos host cusps that remain relevant for annihilation.
We retain $f_{\rm cusp}$ as a free parameter to allow for the
possibility that, in a fraction of halos, prompt cusps become sufficiently
modified by hierarchical evolution that their annihilation contribution is
strongly suppressed.
Exploring $f_{\rm cusp}<1$ therefore provides a conservative way to bracket
uncertainties in the impact of cusp evolution on the annihilation signal.

\subsection{Tidal stripping of hierarchical substructure and cusp survival}
\label{sec:tidal_stripping_cusps}

In SASHIMI, the bound remnant of a subhalo after tidal evolution inside its host
is described by a truncated NFW profile, \textcolor{black}{with the density set to zero at radii greater than the truncation radius $r_t$}.
%such that the density drops rapidly
%outside the truncation radius $r_t$.
%In this effective description, higher-order substructures initially residing at $r>r_t$ are treated as no longer bound to their immediate parent subhalo, and are counted as stripped material in the hierarchical bookkeeping.
%This prescription is intended as an effective treatment of stripping, rather than a detailed model of the phase-space distribution of the stripped debris.}
In this approximation, substructure residing at radii $\textcolor{black}{R}>r_t$ is
interpreted as being stripped from the parent subhalo and incorporated into the
host halo as debris.
\textcolor{black}{A limitation of the present treatment is that SASHIMI describes subhalos
statistically in terms of ensemble-averaged quantities, and does not follow the
individual orbits of subhalos within their hosts. As a result, orbital effects such
as enhanced stripping near pericenter are not modelled explicitly.}

When hierarchical \textcolor{black}{substructures are} included (i.e., $\mathrm{sub}^n$-subhalos),
the same stripping criterion can be applied recursively.
We assume that $\mathrm{sub}^n$-subhalos are spatially distributed within their
immediate host, the $\mathrm{sub}^{n-1}$-subhalo, according to a cored number-density profile \textcolor{black}{(e.g., \cite{Gao:2011rf, Ando:2013ff})},
\begin{equation}
  n_{\rm sh}^{(n)}(\textcolor{black}{R})\ \propto\ \left[\textcolor{black}{R}^2+\bigl(r_s^{(n-1)}\bigr)^2\right]^{-3/2},
\end{equation}
where $r_s^{(n-1)}$ is the scale radius of the $\mathrm{sub}^{n-1}$-subhalo.
\textcolor{black}{Here we denote by $R$ the distance of an $n$th-level subhalo from the center of its
$(n-1)$th-level parent subhalo, reserving $r$ for the internal radial coordinate
used in the density profile of an individual dark matter structure.}
With this prescription, for each $(n\!-\!1)$-level subhalo, only the $n$-level
subhalos located within its truncation radius $r_t^{(n-1)}$ are retained as bound members
of that branch, while those at $\textcolor{black}{R}>r_t$ are classified as stripped.
Because prompt cusps are associated with the internal substructure, this
distinction matters when translating the hierarchical census of subhalos into a
census of {\em surviving} cusps.

We therefore split the effective number of $\mathrm{sub}^n$-subhalos into two
components:
$N_{\rm sh,in}^{(n)}$, corresponding to subhalos located inside the truncation
radius $r_t$ of their immediate parent and thus retained as bound members of the
hierarchy, and
$N_{\rm sh,stripped}^{(n)}$, corresponding to subhalos  outside $r_t$
that are \textcolor{black}{treated as stripped from their immediate parent}.
%released by tidal stripping.
\textcolor{black}{In the hierarchical bookkeeping adopted here, such stripped subhalos are no longer
counted as bound members of that parent, and their mass is regarded as part of the
diffuse component of the higher-level host.}
Physically, the central cusps associated with stripped subhalos are expected to
be resilient against tidal forces.
\textcolor{black}{Accordingly, in our treatment, the stripping of a subhalo and the survival of its
central cusp are treated separately: while the stripped subhalo is absorbed into
the diffuse component of the higher-level host, its cusp may still survive as an
independent annihilation source.}
However, once released into the host halo, the extent to which \textcolor{black}{such a cusp remains}
identifiable as independent contributors to the annihilation signal is uncertain
within a halo-based semi-analytic description.
To bracket this uncertainty, we introduce an effective parameter
$f_{\rm surv,stripped}$, defined as the fraction of prompt cusps associated with
stripped subhalos that contribute independently to the annihilation signal.

For a Milky-Way--size host ($M=10^{12}M_\odot$, $z=0$), the resulting effective
subhalo counts are summarized in Table~\ref{tab:Nsh_hierarchy}.
The number of cusps contributing to the annihilation signal is then written as
\begin{equation}
    N_{\rm cusp}
    =
    f_{\rm cusp}
    \left(
        N_{\rm sh,in}
        +
        f_{\rm surv,stripped}\,
        N_{\rm sh,stripped}
    \right),
    \label{eq:Ncusp}
\end{equation}
where, as discussed in the previous subsection, $f_{\rm cusp}$ controls the overall fraction of halos whose central cusps
contribute effectively to the annihilation signal.
Physically, cusps are expected to survive tidal stripping.
Accordingly, we adopt $f_{\rm surv,stripped}=1$ as the fiducial choice throughout
this work.
\textcolor{black}{In this limit, cusps associated with stripped subhalos are assumed to remain as
independent, effectively naked cusps in the higher-level host, even though their
parent subhalos are no longer counted as bound substructures.}
We nevertheless explore values $f_{\rm surv,stripped}<1$ as a conservative way to
allow for the possibility that cusps associated with stripped substructure
contribute less efficiently to the annihilation signal, for example due to
orbital heating, etc.

\section{Results}
\label{sec:Results}

We now quantify the annihilation enhancement from hierarchical substructure and
prompt cusps in terms of the luminosity parameter,
\begin{equation}
    J \equiv \int d^3\bm{x}\,\rho^2(\bm{x}),
\end{equation}
which directly sets the annihilation rate for a fixed particle-physics model.
The total contribution from prompt cusps can be written as
\begin{equation}
    J_{\rm cusp,\,total}
    = N_{\rm cusp}\!\left(f_{\rm cusp},f_{\rm surv,stripped}\right)\,
      \langle J_{\rm cusp}\rangle,
\end{equation}
where $\langle J_{\rm cusp}\rangle$ is the mean cusp luminosity defined in
Eq.~(\ref{eq:average Jcusp}), and $N_{\rm cusp}(f_{\rm cusp},f_{\rm surv,stripped})$
is given by Eq.~(\ref{eq:Ncusp}).

We express the overall annihilation boost factor as an additive decomposition,
\begin{equation}
    B = B_{\rm sh} + B_{\rm cusp},
\end{equation}
with each component defined by the ratio of the total substructure or cusp
luminosity to that of the smooth host halo,
\begin{equation}
    B_{\rm sh}   \equiv \frac{J_{\rm sh,\,total}}{J_{\rm host}},\qquad
    B_{\rm cusp} \equiv \frac{J_{\rm cusp,\,total}}{J_{\rm host}} .
\end{equation}
For the host, we adopt an NFW density profile characterized by
$(\rho_{s,\rm host}, r_{s,\rm host})$ and concentration $c_{200,\rm host}\equiv r_{200,\rm host}/r_{s,\rm host}$.
The corresponding host luminosity parameter is
\begin{equation}
    J_{\rm host}
    = \int_0^{r_{200}} dr\,4\pi r^2\,\rho_{\rm host}^2(r)
    = \frac{4\pi}{3}\rho_{s,\rm host}^2 r_{s,\rm host}^3
      \left[1-\frac{1}{(1+c_{200,\rm host})^3}\right].
\end{equation}
The total subhalo contribution $J_{\rm sh,\,total}$ is computed as the weighted sum
over the SASHIMI catalog [cf.\ Eq.~(\ref{eq:Jsh})] including the contribution by up to sub$^n$-subhalos, while the cusp component
is obtained from $N_{\rm cusp}$ and $\langle J_{\rm cusp}\rangle$ as above.

\begin{figure}[t]
    \centering
    \includegraphics[width=10cm]{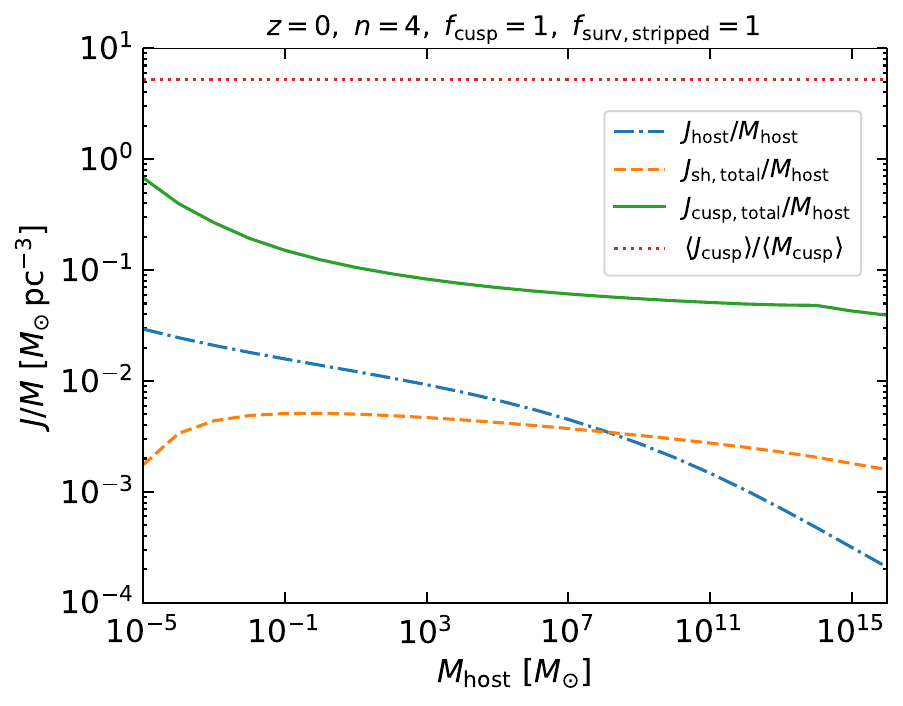}
    \caption{Mass-normalized annihilation luminosity parameters at $z=0$ including hierarchical substructure up to $\mathrm{sub}^4$-subhalos, shown as functions of the host-halo mass $M_{\rm host}$. The dot-dashed, dashed, and solid curves show the contributions from the smooth host-halo component, the total subhalo population, and the total prompt-cusp population, respectively (i.e., $J_{\rm host}/M_{\rm host}$, $J_{\rm sh,,total}/M_{\rm host}$, and $J_{\rm cusp,,total}/M_{\rm host}$), for the fiducial choice of $f_{\rm cusp}=f_{\rm surv,stripped}=1$.
The dotted curve shows the intrinsic cusp annihilation efficiency, $\langle J_{\rm cusp}\rangle/\langle M_{\rm cusp}\rangle$.}
    \label{fig:JperM}
\end{figure}

Figure~\ref{fig:JperM} shows the mass-normalized luminosity parameters that enter the boost-factor
calculation, plotted as functions of the host mass at $z=0$.
The host contribution $J_{\rm host}/M_{\rm host}$ decreases with increasing $M_{\rm host}$,
reflecting the systematic reduction in halo concentration (and hence in $\rho^2$-weighted mass)
toward higher masses.
The subhalo contribution $J_{\rm sh,\,total}/M_{\rm host}$ varies more gently with $M_{\rm host}$,
as it represents an integral over the subhalo population whose internal structure and abundance
evolve only mildly across the mass range considered.
In contrast, the prompt-cusp contribution $J_{\rm cusp,\,total}/M_{\rm host}$ remains comparatively
large over a broad range of host masses for our fiducial parameters ($f_{\rm cusp} = f_{\rm surv,stripped}=1$), illustrating that the cusps can yield a dominant contribution to the total annihilation budget.

For comparison, we also show $\langle J_{\rm cusp}\rangle/\langle M_{\rm cusp}\rangle$---the mean cusp luminosity per unit cusp mass,
which is host-mass independent and is primarily controlled by the
cusp internal structure and the underlying microphysical inputs.
We note that $\langle J_{\rm cusp}\rangle/\langle M_{\rm cusp}\rangle$ corresponds to the limiting case in which the annihilation luminosity is entirely generated by
cusp-like material, i.e.\ as if all dark matter were converted into cusps with the same internal
structure. It therefore provides a useful absolute reference scale and can be viewed as an
upper-envelope benchmark for the $J_{\rm cusp,\,total}/M_{\rm host}$ curve, which is necessarily
suppressed by the fact that only a fraction of the host mass is bound into surviving cusps in the
hierarchical assembly.

\begin{figure}[t]
    \centering
    \includegraphics[width=10cm]{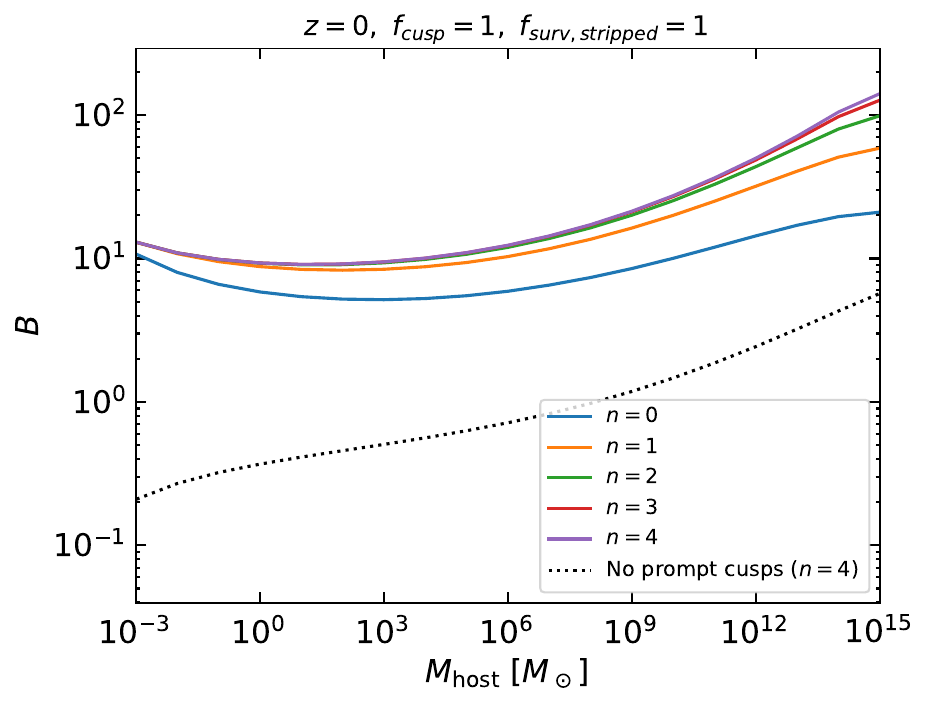}
    \caption{
    Dependence of the total boost factor $B$ on the depth of hierarchical substructure included in the model.
    Curves show results at $z=0$ for $n=0,\ldots,4$ (from bottom to top), where $n$ denotes inclusion up to $\mathrm{sub}^n$-subhalos.
    The dotted curve shows the corresponding reference prediction without prompt cusps (for $n=4$).
    }
    \label{fig:Bhier}
\end{figure}

Figure~\ref{fig:Bhier} shows the resulting annihilation boost factor at $z=0$ for the
converged hierarchical model, adopting the fiducial parameter choice
$f_{\rm cusp}=1$ and $f_{\rm surv,stripped}=1$.
Relative to the standard subhalo-only prediction (dotted curve), the inclusion
of prompt cusps enhances the boost \textcolor{black}{substantially}.
It also illustrates how the predicted annihilation boost changes
as progressively deeper levels of hierarchical substructure are included.
The sequence from $n=0$ to $n=4$ shows that the boost rapidly approaches a
converged result once a few substructure levels are accounted for, indicating
that the hierarchical expansion is well controlled.
This figure therefore summarizes the main physical impact of prompt cusps in a
fully hierarchical setting.

\begin{figure}[t]
    \centering
    \includegraphics[width=10cm]{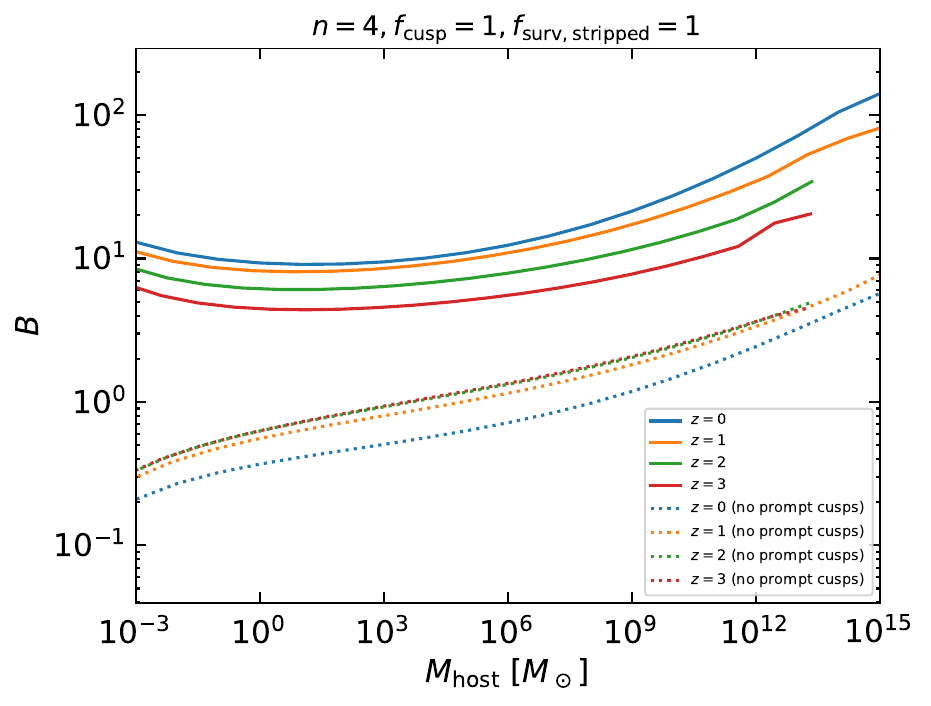}
    \caption{
    Redshift dependence of the total boost factor $B$ as a function of host-halo mass $M_{\rm host}$.
    Solid curves include prompt cusps, while dotted curves show the corresponding no-cusp baseline,
    for $z=0,1,2$, and 3 (from top to bottom at high masses).  
    }
    \label{fig:Bz}
\end{figure}

Figure~\ref{fig:Bz} shows the redshift evolution of the boost factor for the same
fiducial model.
\textcolor{black}{At fixed host mass, the boost varies with redshift for several reasons.
Firstly, the structural properties of halos evolve with redshift.
Secondly, at earlier epochs there has been less time for hierarchical buildup, so
the host contains a less developed population of nested substructure.
Thirdly, substructures that have already formed have undergone less tidal
processing, since they have had less time to be stripped or disrupted after
accretion.
In all cases, the inclusion of prompt cusps shifts the boost above the no-cusp
baseline while preserving the overall mass dependence.}
%At fixed host mass, the boost varies with redshift due to the evolving structural
%properties of halos and the reduced time available for hierarchical buildup and
%tidal processing at earlier epochs.
%In all cases, the inclusion of prompt cusps shifts the boost above the no-cusp
%baseline while preserving the overall mass dependence.

\begin{figure}[t]
    \centering
    \includegraphics[width=10cm]{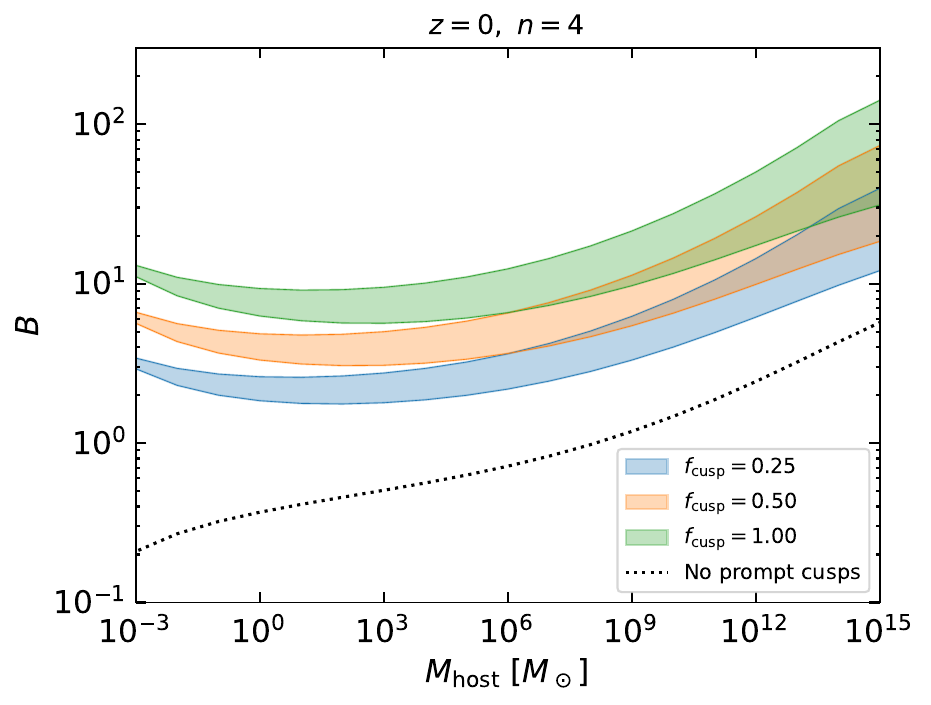} % <-- replace with your filename
    \caption{Annihilation boost factor $B = B_{\rm sh}+B_{\rm cusp}$ at $z=0$ including hierarchical substructure up to $\mathrm{sub}^4$-subhalos, shown as a function of the host-halo mass $M_{\rm host}$.
    The colored bands correspond to different choices of the cusp-occupation fraction $f_{\rm cusp}$ (0.25, 0.5, and 1 from bottom to top).
    For each $f_{\rm cusp}$, the band width reflects uncertainty in the survival of cusps associated with the stripped region, parametrized by $f_{\rm surv,stripped}\in[0,1]$:
    the lower (upper) edge assumes $f_{\rm surv,stripped}=0$ ($1$).
    The dotted curve shows the baseline prediction without prompt cusps ($B = B_{\rm sh}$).}
    \label{fig:B}
\end{figure}

Finally, we assess the sensitivity of our results to the modelling assumptions
associated with prompt cusps.
In Fig.~\ref{fig:B}, the colored bands indicate the range obtained by varying the
cusp-occupation fraction $f_{\rm cusp}$ and the stripped-cusp contribution
parameter $f_{\rm surv,stripped}$ around their fiducial values.
The overall normalization of the boost is controlled by $f_{\rm cusp}$, while
the band width reflects the additional uncertainty associated with cusps
originating from stripped substructure.

\section{Discussion}
\label{sec:Discussion}

\subsection{Comparison with earlier work}

For similar benchmark assumptions as in the universal-average treatment of
Ref.~\cite{delos2023prompt} ($f_{\rm cusp}=f_{\rm surv,stripped}=1$),
our predicted Milky-Way--size boost is $B\sim 50$ at $z=0$,
about a factor of four smaller than the $\sim 200$ quoted by Delos and White~\cite{delos2023prompt}.
This offset can be understood primarily as a difference in the inferred abundance
of surviving cusps: they have $N_{\rm cusp}\sim 2\times 10^{16}$ in a
$10^{12}M_\odot$ halo, whereas Table~\ref{tab:Nsh_hierarchy} combined with
Eq.~(\ref{eq:Ncusp}) yields $N_{\rm cusp}\sim 10^{16}$.
Since $J_{\rm cusp,total}\propto N_{\rm cusp}\langle J_{\rm cusp}\rangle$ at fixed cusp
internal structure, this difference in $N_{\rm cusp}$ maps directly onto that in $B$.
Equivalently, adopting the typical prompt-cusp mass $\langle M_{\rm cusp}\rangle\sim 10^{-6}M_\odot$
(Eq.~\ref{eq:average Jcusp} and surrounding discussion), the implied total mass bound in cusps is
$M_{\rm cusp,tot}\sim N_{\rm cusp}\langle M_{\rm cusp}\rangle \sim 2\times 10^{10}M_\odot$
in the Delos and White estimate (i.e.\ $\sim 2\%$ of a $10^{12}M_\odot$ host),
while our hierarchical \textsc{SASHIMI}-based count implies
$M_{\rm cusp,tot}\sim 10^{10}M_\odot$ (i.e.\ $\sim 1\%$ of the host mass).

\subsection{Possible origin of the cusp-abundance difference}

At present, the origin of the cusp-count mismatch can only be discussed qualitatively,
and a quantitative reconciliation is beyond the scope of this work.
Nevertheless, several differences in modeling philosophy plausibly contribute.

First, our cusp abundance inherits the EPS-based merger-tree construction underlying
\textsc{SASHIMI}, whereas Ref.~\cite{delos2023prompt} builds cusp statistics from peak
theory using a universal-average mapping.
EPS prescriptions are often calibrated primarily on relatively massive haloes, so their
extrapolation into the microhalo regime is not guaranteed \emph{a priori}.
On the other hand, recent ultra-high-resolution simulations of void environments find that
EPS, treated as a conditional mass function, can remain accurate down to $M\sim 10^{-6}M_\odot$
in underdense regions~\cite{Zheng:2023myp}.
This supports the plausibility of EPS-based extrapolations in at least some environments,
while leaving open the possibility that peak-selection effects or environmental conditioning
can generate larger deviations in massive hosts.

%Second, Ref.~\cite{delos2023prompt} effectively counts cusps associated with peaks that
%collapse down to $z=0$.
%In contrast, our implementation assigns cusps at the formation of (sub)halos within the
%hierarchical tree and does not create additional cusps within an already-accreted branch;
%subsequent evolution can only reduce the identifiable cusp population through stripping or
%disruption.
%If late-time cusp formation in already-accreted material is non-negligible, our procedure
%will undercount cusps relative to a universal peak-based estimate.

Second, microhalo-scale structures that host primordial cusps are expected to be weakly
clustered ($b<1$) relative to dark matter particles in many environments, whereas a universal-average
mapping effectively assumes an unbiased distribution when translating ``cusps per unit mass''
into ``cusps per host'' independent of environment.
Even a modest anti-bias could reduce the expected cusp abundance in Milky-Way--like regions
relative to the universal estimate.

Lastly, an additional, and potentially important, source of uncertainty in the EPS-based
microhalo abundance concerns the mapping between the cutoff scale and halo mass when
computing $\sigma(M)$ for a suppressed small-scale power spectrum.
In \textsc{SASHIMI} we evaluate $\sigma(M)$ with a sharp-$k$ filter and adopt a cutoff
$k_{\max}=\alpha/R$, where $\alpha$ parametrizes the residual ambiguity of this mapping
\cite{Schneider:2013ria}.
Our fiducial choice, $\alpha=1.8$, is motivated by matching to the real-space top-hat
result in the large-mass regime, but it is not calibrated directly on microhalo
simulations.
Crucially, varying $\alpha$ can change the predicted abundance of microhaloes---and hence
the inferred number of prompt cusps---without dramatically affecting the mass function
on the high-mass end that dominates typical calibrations.
For example, we find that choosing $\alpha=1$ increases the predicted cusp abundance by
a factor of $\sim 5$ relative to the fiducial model, while $\alpha=2.5$ reduces it to
$\sim 0.4$ times the fiducial value.
This sensitivity suggests that a dedicated calibration of $\alpha$ in the microhalo regime
(e.g.\ using ultra-high-resolution simulations that resolve Earth-mass haloes and their
environmental dependence) will be important for robust predictions of $N_{\rm cusp}$ and,
in turn, of the cusp-induced annihilation boost.

\subsection{\textcolor{black}{Constraints from gamma-ray observations}}

\textcolor{black}{A full update of the gamma-ray phenomenology, as was discussed in
Refs.~\cite{delos2024limits, Crnogorcevic:2025nwp}, including constraints from Galactic, cluster, and
especially diffuse extragalactic signals, is beyond the scope of the present work.
For individual host halos, the main effect of our revised calculation would be a
reduction of the predicted signal in line with the smaller boost factor obtained
here compared with Ref.~\cite{delos2023prompt}. By contrast, updating the diffuse extragalactic constraints requires a
dedicated cosmological calculation including the contribution of low-mass halos over
a wide range of masses and redshifts, which we leave for future work.}

\subsection{Future directions}

A definitive resolution of the difference to universal-average predictions
likely requires a unified framework that embeds peak-based cusp formation criteria into
merger-tree evolution with explicit environmental conditioning.
There is relevant groundwork connecting peaks and excursion-set ideas, such as excursion-set
peaks and related peak-constrained excursion-set formalisms \cite{Paranjape:2012ks,Paranjape:2012jt}.
Complementary peak-based methods that generate halo catalogs and assembly histories directly
from the initial density field---e.g.\ peak-patch \cite{Bond:1993we} and PINOCCHIO
\cite{Monaco:2001jg,Taffoni:2001jh}---may also provide useful routes for marrying
cusp-scale peak properties to hierarchical growth.
Adapting such approaches to prompt cusps (including the mapping from peak parameters to cusp
structural parameters and survival probabilities) is an interesting direction for future work.

\section{Conclusions}
\label{sec:Conclusions}

Recent simulations indicate that the earliest collapsing peaks in the primordial
density field can form steep ``prompt cusps'' with $\rho\propto r^{-3/2}$ and that a
substantial fraction may survive hierarchical growth to $z=0$ in WIMP-like CDM scenarios
\cite{10.1093/mnras/stac3373,delos2023prompt,delos2024limits}.
Motivated by these developments, we implemented prompt cusps in the semi-analytic
substructure framework \textsc{SASHIMI}~\cite{Hiroshima:2018kfv,Ando_2019}, enabling a
hierarchical and environment-dependent treatment of annihilation enhancement beyond the
universal-average boost approach.
We summarize our main findings as follows:
\begin{itemize}
\item The predicted contribution from hierarchical substructure converges rapidly with the depth
of the hierarchy: including up to $\mathrm{sub}^3$--$\mathrm{sub}^4$-subhalos is sufficient for
a stable boost prediction in the mass range studied (Fig.~\ref{fig:Bhier}).

\item Prompt cusps provide an additional component to the annihilation luminosity that can raise
the total boost above the standard subhalo-only expectation.
For fiducial survival assumptions ($f_{\rm cusp}=1$ and $f_{\rm surv,stripped}=1$), we find
$B\sim 50$ for Milky-Way--like hosts at $z=0$, while the no-cusp
baseline remains at $B_{\rm sh}\sim\mathrm{few}$ (Fig.~\ref{fig:Bhier}).

\item Across host masses and redshifts, prompt cusps increase the normalization of $B(M_{\rm host},z)$
while largely preserving the qualitative mass and redshift trends of the baseline prediction (Fig.~\ref{fig:Bz}).
Our predictions are based on the fiducial assumption that prompt cusps are
present in all collapsed halos and remain relevant for annihilation.
The robustness of this result against modelling uncertainties is assessed by
varying the cusp-occupation fraction $f_{\rm cusp}$ and the stripped-cusp
contribution parameter $f_{\rm surv,stripped}$, as illustrated in
Fig.~\ref{fig:B}.

\end{itemize}

Compared to the universal-average peak-based estimate of Delos and White~\cite{delos2023prompt}, our
fiducial Milky-Way boost is lower by about a factor of four, which we attribute primarily to a
correspondingly lower inferred number of cusps in a host.
This difference highlights the importance of unifying peak-based cusp formation statistics with
merger-tree evolution and environmental conditioning.
Our \textsc{SASHIMI}-based implementation provides a flexible platform for pursuing this goal and for
propagating improved cusp-formation and survival prescriptions into testable predictions for annihilation
signals.

%\appendix
%\section{Some title}
%Please always give a title also for appendices.

\acknowledgments

We thank M.~Sten Delos for helpful discussion. This work was partly supported by MEXT KAKENHI Grant Numbers, JP20H05850, JP20H05861, and JP24K07039.

%\paragraph{Note added.} This is also a good position for notes added
%after the paper has been written.

% Bibliography

%% [A] Recommended: using JHEP.bst file
\bibliographystyle{JHEP}
\bibliography{references.bib}

@article{Hofmann_2001,
   title={Damping scales of neutralino cold dark matter},
   volume={64},
   ISSN={1089-4918},
   url={http://dx.doi.org/10.1103/PhysRevD.64.083507},
   DOI={10.1103/physrevd.64.083507},
   number={8},
   journal={Physical Review D},
   publisher={American Physical Society (APS)},
   author={Hofmann, Stefan and Schwarz, Dominik J. and Stöcker, Horst},
   year={2001},
   month=sep }

@article{PhysRevLett.97.031301,
  title = {What Mass Are the Smallest Protohalos?},
  author = {Profumo, Stefano and Sigurdson, Kris and Kamionkowski, Marc},
  journal = {Phys. Rev. Lett.},
  volume = {97},
  issue = {3},
  pages = {031301},
  numpages = {4},
  year = {2006},
  month = {Jul},
  publisher = {American Physical Society},
  doi = {10.1103/PhysRevLett.97.031301},
  url = {https://link.aps.org/doi/10.1103/PhysRevLett.97.031301}
}

@article{Navarro_1997,
doi = {10.1086/304888},
url = {https://dx.doi.org/10.1086/304888},
year = {1997},
month = {dec},
publisher = {},
volume = {490},
number = {2},
pages = {493},
author = {Julio F. Navarro and Carlos S. Frenk and Simon D. M. White},
title = {A Universal Density Profile from Hierarchical Clustering},
journal = {The Astrophysical Journal}
}

@article{Ando_2019,
   title={Halo Substructure Boosts to the Signatures of Dark Matter Annihilation},
   volume={7},
   ISSN={2075-4434},
   url={http://dx.doi.org/10.3390/galaxies7030068},
   DOI={10.3390/galaxies7030068},
   number={3},
   journal={Galaxies},
   publisher={MDPI AG},
   author={Ando, Shin’ichiro and Ishiyama, Tomoaki and Hiroshima, Nagisa},
   year={2019},
   month=jul, pages={68} }

@article{Loeb:2005pm,
    author = "Loeb, Abraham and Zaldarriaga, Matias",
    title = "{The Small-scale power spectrum of cold dark matter}",
    eprint = "astro-ph/0504112",
    archivePrefix = "arXiv",
    doi = "10.1103/PhysRevD.71.103520",
    journal = "Phys. Rev. D",
    volume = "71",
    pages = "103520",
    year = "2005"
}

@article{Bertschinger:2006nq,
    author = "Bertschinger, Edmund",
    title = "{The Effects of Cold Dark Matter Decoupling and Pair Annihilation on Cosmological Perturbations}",
    eprint = "astro-ph/0607319",
    archivePrefix = "arXiv",
    doi = "10.1103/PhysRevD.74.063509",
    journal = "Phys. Rev. D",
    volume = "74",
    pages = "063509",
    year = "2006"
}

@misc{delos2024limits,
      title={Limits on dark matter annihilation in prompt cusps from the isotropic gamma-ray background}, 
      author={M. Sten Delos and Michael Korsmeier and Axel Widmark and Carlos Blanco and Tim Linden and Simon D. M. White},
      year={2024},
      eprint={2307.13023},
      archivePrefix={arXiv},
      primaryClass={astro-ph.HE}
}

@article{Crnogorcevic:2025nwp,
    author = "Crnogor{\v{c}}evi{\'c}, Milena and Delos, M. Sten and Kuritz{\'e}n, Nadia and Linden, Tim",
    title = "{Gamma-ray observations of galaxy clusters strongly constrain dark matter annihilation in prompt cusps}",
    eprint = "2501.14865",
    archivePrefix = "arXiv",
    primaryClass = "astro-ph.HE",
    doi = "10.1103/wtkl-nkdw",
    journal = "Phys. Rev. D",
    volume = "112",
    number = "10",
    pages = "103001",
    year = "2025"
}

@article{Diamanti:2015kma,
    author = "Diamanti, Roberta and Catalan, Maria Eugenia Cabrera and Ando, Shin'ichiro",
    title = "{Dark matter protohalos in a nine parameter MSSM and implications for direct and indirect detection}",
    eprint = "1506.01529",
    archivePrefix = "arXiv",
    primaryClass = "hep-ph",
    doi = "10.1103/PhysRevD.92.065029",
    journal = "Phys. Rev. D",
    volume = "92",
    number = "6",
    pages = "065029",
    year = "2015"
}

@article{delos2023prompt,
    author = "Delos, M. Sten and White, Simon D. M.",
    title = "{Prompt cusps and the dark matter annihilation signal}",
    eprint = "2209.11237",
    archivePrefix = "arXiv",
    primaryClass = "astro-ph.CO",
    doi = "10.1088/1475-7516/2023/10/008",
    journal = "JCAP",
    volume = "10",
    pages = "008",
    year = "2023"
}

@article{Ishiyama:2010es,
    author = "Ishiyama, Tomoaki and Makino, Junichiro and Ebisuzaki, Toshikazu",
    title = "{Gamma-ray Signal from Earth-mass Dark Matter Microhalos}",
    eprint = "1006.3392",
    archivePrefix = "arXiv",
    primaryClass = "astro-ph.CO",
    doi = "10.1088/2041-8205/723/2/L195",
    journal = "Astrophys. J. Lett.",
    volume = "723",
    pages = "L195",
    year = "2010"
}

@article{Anderhalden:2013wd,
    author = "Anderhalden, Donnino and Diemand, Juerg",
    title = "{Density Profiles of CDM Microhalos and their Implications for Annihilation Boost Factors}",
    eprint = "1302.0003",
    archivePrefix = "arXiv",
    primaryClass = "astro-ph.CO",
    doi = "10.1088/1475-7516/2013/04/009",
    journal = "JCAP",
    volume = "04",
    pages = "009",
    year = "2013",
    note = "[Erratum: JCAP 08, E02 (2013)]"
}

@article{Ishiyama:2014uoa,
    author = "Ishiyama, Tomoaki",
    title = "{Hierarchical Formation of Dark Matter Halos and the Free Streaming Scale}",
    eprint = "1404.1650",
    archivePrefix = "arXiv",
    primaryClass = "astro-ph.CO",
    doi = "10.1088/0004-637X/788/1/27",
    journal = "Astrophys. J.",
    volume = "788",
    pages = "27",
    year = "2014"
}

@article{10.1093/mnras/stac3373,
    author = {Delos, M Sten and White, Simon D M},
    title = "{Inner cusps of the first dark matter haloes: formation and survival in a cosmological context}",
    journal = {Monthly Notices of the Royal Astronomical Society},
    volume = {518},
    number = {3},
    pages = {3509-3532},
    year = {2022},
    month = {11},
    issn = {0035-8711},
    doi = {10.1093/mnras/stac3373},
    url = {https://doi.org/10.1093/mnras/stac3373},
    eprint = {https://academic.oup.com/mnras/article-pdf/518/3/3509/47466116/stac3373.pdf},
}

@article{Green:2003un,
    author = "Green, Anne M. and Hofmann, Stefan and Schwarz, Dominik J.",
    title = "{The power spectrum of SUSY - CDM on sub-galactic scales}",
    eprint = "astro-ph/0309621",
    archivePrefix = "arXiv",
    doi = "10.1111/j.1365-2966.2004.08232.x",
    journal = "Mon. Not. Roy. Astron. Soc.",
    volume = "353",
    pages = "L23",
    year = "2004"
}

@article{Tremaine:1979we,
    author = "Tremaine, S. and Gunn, J. E.",
    editor = "Srednicki, M. A.",
    title = "{Dynamical Role of Light Neutral Leptons in Cosmology}",
    doi = "10.1103/PhysRevLett.42.407",
    journal = "Phys. Rev. Lett.",
    volume = "42",
    pages = "407--410",
    year = "1979"
}

@article{Silk:1992bh,
    author = "Silk, Joseph and Stebbins, Albert",
    title = "{Clumpy cold dark matter}",
    reportNumber = "CFPA-TH-92-09, FERMILAB-PUB-93-031-A",
    doi = "10.1086/172846",
    journal = "Astrophys. J.",
    volume = "411",
    pages = "439--449",
    year = "1993"
}

@article{Bergstrom:1998jj,
    author = "Bergstrom, Lars and Edsjo, Joakim and Gondolo, Paolo and Ullio, Piero",
    title = "{Clumpy neutralino dark matter}",
    eprint = "astro-ph/9806072",
    archivePrefix = "arXiv",
    reportNumber = "USITP-98-08, MPI-PHT-98-43",
    doi = "10.1103/PhysRevD.59.043506",
    journal = "Phys. Rev. D",
    volume = "59",
    pages = "043506",
    year = "1999"
}

@article{Bergstrom:1998zs,
    author = "Bergstrom, Lars and Edsjo, Joakim and Ullio, Piero",
    title = "{Possible indications of a clumpy dark matter halo}",
    eprint = "astro-ph/9804050",
    archivePrefix = "arXiv",
    doi = "10.1103/PhysRevD.58.083507",
    journal = "Phys. Rev. D",
    volume = "58",
    pages = "083507",
    year = "1998"
}

@article{Stoehr:2003hf,
    author = "Stoehr, Felix and White, Simon D. M. and Springel, Volker and Tormen, Giuseppe and Yoshida, Naoki",
    title = "{Dark matter annihilation in the halo of the Milky Way}",
    eprint = "astro-ph/0307026",
    archivePrefix = "arXiv",
    doi = "10.1046/j.1365-2966.2003.07052.x",
    journal = "Mon. Not. Roy. Astron. Soc.",
    volume = "345",
    pages = "1313",
    year = "2003"
}

@article{Koushiappas:2003bn,
    author = "Koushiappas, Savvas M. and Zentner, Andrew R. and Walker, Terrence P.",
    title = "{The observability of gamma-rays from neutralino annihilations in Milky Way substructure}",
    eprint = "astro-ph/0309464",
    archivePrefix = "arXiv",
    doi = "10.1103/PhysRevD.69.043501",
    journal = "Phys. Rev. D",
    volume = "69",
    pages = "043501",
    year = "2004"
}

@article{Ando:2005hr,
    author = "Ando, Shin'ichiro",
    title = "{Can dark matter annihilation dominate the extragalactic gamma-ray background?}",
    eprint = "astro-ph/0503006",
    archivePrefix = "arXiv",
    reportNumber = "UTAP-513",
    doi = "10.1103/PhysRevLett.94.171303",
    journal = "Phys. Rev. Lett.",
    volume = "94",
    pages = "171303",
    year = "2005"
}

@article{Pieri:2007ir,
    author = "Pieri, L. and Bertone, G. and Branchini, E",
    title = "{Dark Matter Annihilation in Substructures Revised}",
    eprint = "0706.2101",
    archivePrefix = "arXiv",
    primaryClass = "astro-ph",
    doi = "10.1111/j.1365-2966.2007.12828.x",
    journal = "Mon. Not. Roy. Astron. Soc.",
    volume = "384",
    pages = "1627",
    year = "2008"
}

@article{Berezinsky:2006qm,
    author = "Berezinsky, Veniamin and Dokuchaev, Vyacheslav and Eroshenko, Yury",
    title = "{Anisotropy of dark matter annihilation with respect to the Galactic plane}",
    eprint = "astro-ph/0612733",
    archivePrefix = "arXiv",
    doi = "10.1088/1475-7516/2007/07/011",
    journal = "JCAP",
    volume = "07",
    pages = "011",
    year = "2007"
}

@article{Lavalle:2007apj,
    author = "Lavalle, J. and Yuan, Q. and Maurin, D. and Bi, X. J.",
    title = "{Full Calculation of Clumpiness Boost factors for Antimatter Cosmic Rays in the light of Lambda-CDM N-body simulation results. Abandoning hope in clumpiness enhancement?}",
    eprint = "0709.3634",
    archivePrefix = "arXiv",
    primaryClass = "astro-ph",
    reportNumber = "CPPM-P-2007-02",
    doi = "10.1051/0004-6361:20078723",
    journal = "Astron. Astrophys.",
    volume = "479",
    pages = "427--452",
    year = "2008"
}

@article{stucker2023effect,
    author = {St{\"u}cker, Jens and Ogiya, Go and White, Simon D. M. and Angulo, Raul E.},
    title = "{The effect of stellar encounters on the dark matter annihilation signal from prompt cusps}",
    eprint = "2301.04670",
    archivePrefix = "arXiv",
    primaryClass = "astro-ph.CO",
    doi = "10.1093/mnras/stad1268",
    journal = "Mon. Not. Roy. Astron. Soc.",
    volume = "523",
    number = "1",
    pages = "1067--1088",
    year = "2023"
}

@article{Bardeen:1985tr,
    author = "Bardeen, James M. and Bond, J. R. and Kaiser, Nick and Szalay, A. S.",
    title = "{The Statistics of Peaks of Gaussian Random Fields}",
    reportNumber = "FERMILAB-PUB-85-148-A, NSF-ITP-85-93",
    doi = "10.1086/164143",
    journal = "Astrophys. J.",
    volume = "304",
    pages = "15--61",
    year = "1986"
}

@article{Eisenstein:1997ik,
    author = "Eisenstein, Daniel J. and Hu, Wayne",
    title = "{Baryonic features in the matter transfer function}",
    eprint = "astro-ph/9709112",
    archivePrefix = "arXiv",
    reportNumber = "IASSNS-AST-97-51",
    doi = "10.1086/305424",
    journal = "Astrophys. J.",
    volume = "496",
    pages = "605",
    year = "1998"
}

@article{PhysRevD.100.023523,
  title = {Predicting the density profiles of the first halos},
  author = {Delos, M. Sten and Bruff, Margie and Erickcek, Adrienne L.},
  journal = {Phys. Rev. D},
  volume = {100},
  issue = {2},
  pages = {023523},
  numpages = {25},
  year = {2019},
  month = {Jul},
  publisher = {American Physical Society},
  doi = {10.1103/PhysRevD.100.023523},
  url = {https://link.aps.org/doi/10.1103/PhysRevD.100.023523}
}

@article{refId0,
	author = {{Planck Collaboration} and {Aghanim, N.} and {Akrami, Y.} and {Ashdown, M.} and {Aumont, J.} and {Baccigalupi, C.} and {Ballardini, M.} and {Banday, A. J.} and {Barreiro, R. B.} and {Bartolo, N.} and {Basak, S.} and {Battye, R.} and {Benabed, K.} and {Bernard, J.-P.} and {Bersanelli, M.} and {Bielewicz, P.} and {Bock, J. J.} and {Bond, J. R.} and {Borrill, J.} and {Bouchet, F. R.} and {Boulanger, F.} and {Bucher, M.} and {Burigana, C.} and {Butler, R. C.} and {Calabrese, E.} and {Cardoso, J.-F.} and {Carron, J.} and {Challinor, A.} and {Chiang, H. C.} and {Chluba, J.} and {Colombo, L. P. L.} and {Combet, C.} and {Contreras, D.} and {Crill, B. P.} and {Cuttaia, F.} and {de Bernardis, P.} and {de Zotti, G.} and {Delabrouille, J.} and {Delouis, J.-M.} and {Di Valentino, E.} and {Diego, J. M.} and {Doré, O.} and {Douspis, M.} and {Ducout, A.} and {Dupac, X.} and {Dusini, S.} and {Efstathiou, G.} and {Elsner, F.} and {Enßlin, T. A.} and {Eriksen, H. K.} and {Fantaye, Y.} and {Farhang, M.} and {Fergusson, J.} and {Fernandez-Cobos, R.} and {Finelli, F.} and {Forastieri, F.} and {Frailis, M.} and {Fraisse, A. A.} and {Franceschi, E.} and {Frolov, A.} and {Galeotta, S.} and {Galli, S.} and {Ganga, K.} and {Génova-Santos, R. T.} and {Gerbino, M.} and {Ghosh, T.} and {González-Nuevo, J.} and {Górski, K. M.} and {Gratton, S.} and {Gruppuso, A.} and {Gudmundsson, J. E.} and {Hamann, J.} and {Handley, W.} and {Hansen, F. K.} and {Herranz, D.} and {Hildebrandt, S. R.} and {Hivon, E.} and {Huang, Z.} and {Jaffe, A. H.} and {Jones, W. C.} and {Karakci, A.} and {Keihänen, E.} and {Keskitalo, R.} and {Kiiveri, K.} and {Kim, J.} and {Kisner, T. S.} and {Knox, L.} and {Krachmalnicoff, N.} and {Kunz, M.} and {Kurki-Suonio, H.} and {Lagache, G.} and {Lamarre, J.-M.} and {Lasenby, A.} and {Lattanzi, M.} and {Lawrence, C. R.} and {Le Jeune, M.} and {Lemos, P.} and {Lesgourgues, J.} and {Levrier, F.} and {Lewis, A.} and {Liguori, M.} and {Lilje, P. B.} and {Lilley, M.} and {Lindholm, V.} and {López-Caniego, M.} and {Lubin, P. M.} and {Ma, Y.-Z.} and {Macías-Pérez, J. F.} and {Maggio, G.} and {Maino, D.} and {Mandolesi, N.} and {Mangilli, A.} and {Marcos-Caballero, A.} and {Maris, M.} and {Martin, P. G.} and {Martinelli, M.} and {Martínez-González, E.} and {Matarrese, S.} and {Mauri, N.} and {McEwen, J. D.} and {Meinhold, P. R.} and {Melchiorri, A.} and {Mennella, A.} and {Migliaccio, M.} and {Millea, M.} and {Mitra, S.} and {Miville-Deschênes, M.-A.} and {Molinari, D.} and {Montier, L.} and {Morgante, G.} and {Moss, A.} and {Natoli, P.} and {Nørgaard-Nielsen, H. U.} and {Pagano, L.} and {Paoletti, D.} and {Partridge, B.} and {Patanchon, G.} and {Peiris, H. V.} and {Perrotta, F.} and {Pettorino, V.} and {Piacentini, F.} and {Polastri, L.} and {Polenta, G.} and {Puget, J.-L.} and {Rachen, J. P.} and {Reinecke, M.} and {Remazeilles, M.} and {Renzi, A.} and {Rocha, G.} and {Rosset, C.} and {Roudier, G.} and {Rubiño-Martín, J. A.} and {Ruiz-Granados, B.} and {Salvati, L.} and {Sandri, M.} and {Savelainen, M.} and {Scott, D.} and {Shellard, E. P. S.} and {Sirignano, C.} and {Sirri, G.} and {Spencer, L. D.} and {Sunyaev, R.} and {Suur-Uski, A.-S.} and {Tauber, J. A.} and {Tavagnacco, D.} and {Tenti, M.} and {Toffolatti, L.} and {Tomasi, M.} and {Trombetti, T.} and {Valenziano, L.} and {Valiviita, J.} and {Van Tent, B.} and {Vibert, L.} and {Vielva, P.} and {Villa, F.} and {Vittorio, N.} and {Wandelt, B. D.} and {Wehus, I. K.} and {White, M.} and {White, S. D. M.} and {Zacchei, A.} and {Zonca, A.}},
	title = {Planck 2018 results - VI. Cosmological parameters},
	DOI= "10.1051/0004-6361/201833910",
	url= "https://doi.org/10.1051/0004-6361/201833910",
	journal = {Astron. Astrophys.},
	year = 2020,
	volume = 641,
	pages = "A6",
}

@article{Sheth:1999su,
    author = "Sheth, Ravi K. and Mo, H. J. and Tormen, Giuseppe",
    title = "{Ellipsoidal collapse and an improved model for the number and spatial distribution of dark matter haloes}",
    eprint = "astro-ph/9907024",
    archivePrefix = "arXiv",
    doi = "10.1046/j.1365-8711.2001.04006.x",
    journal = "Mon. Not. Roy. Astron. Soc.",
    volume = "323",
    pages = "1",
    year = "2001"
}

@article{Bartels:2015uba,
    author = "Bartels, Richard and Ando, Shin'ichiro",
    title = "{Boosting the annihilation boost: Tidal effects on dark matter subhalos and consistent luminosity modeling}",
    eprint = "1507.08656",
    archivePrefix = "arXiv",
    primaryClass = "astro-ph.CO",
    doi = "10.1103/PhysRevD.92.123508",
    journal = "Phys. Rev. D",
    volume = "92",
    number = "12",
    pages = "123508",
    year = "2015"
}

@article{Ishiyama:2019hmh,
    author = "Ishiyama, Tomoaki and Ando, Shin'ichiro",
    title = "{The Abundance and Structure of Subhaloes near the Free Streaming Scale and Their Impact on Indirect Dark Matter Searches}",
    eprint = "1907.03642",
    archivePrefix = "arXiv",
    primaryClass = "astro-ph.CO",
    doi = "10.1093/mnras/staa069",
    journal = "Mon. Not. Roy. Astron. Soc.",
    volume = "492",
    number = "3",
    pages = "3662--3671",
    year = "2020"
}

@article{Hiroshima:2018kfv,
    author = "Hiroshima, Nagisa and Ando, Shin'ichiro and Ishiyama, Tomoaki",
    title = "{Modeling evolution of dark matter substructure and annihilation boost}",
    eprint = "1803.07691",
    archivePrefix = "arXiv",
    primaryClass = "astro-ph.CO",
    reportNumber = "KEK-TH-2043",
    doi = "10.1103/PhysRevD.97.123002",
    journal = "Phys. Rev. D",
    volume = "97",
    number = "12",
    pages = "123002",
    year = "2018"
}

@article{Schneider:2013ria,
    author = "Schneider, Aurel and Smith, Robert E. and Reed, Darren",
    title = "{Halo Mass Function and the Free Streaming Scale}",
    eprint = "1303.0839",
    archivePrefix = "arXiv",
    primaryClass = "astro-ph.CO",
    doi = "10.1093/mnras/stt829",
    journal = "Mon. Not. Roy. Astron. Soc.",
    volume = "433",
    pages = "1573",
    year = "2013"
}

@article{Bond:1993we,
    author = "Bond, J. R. and Myers, S. T.",
    title = "{The Hierarchical peak patch picture of cosmic catalogs. 1. Algorithms}",
    reportNumber = "CITA-93-27",
    doi = "10.1086/192267",
    journal = "Astrophys. J. Suppl.",
    volume = "103",
    pages = "1",
    year = "1996"
}

@article{Bond:1993wd,
    author = "Bond, J. R. and Myers, S. T.",
    title = "{The Hierarchical peak patch picture of cosmic catalogs. 2. Validation and application to clusters}",
    reportNumber = "CITA-93-28",
    doi = "10.1086/192268",
    journal = "Astrophys. J. Suppl.",
    volume = "103",
    pages = "41",
    year = "1996"
}

@article{Zheng:2023myp,
    author = "Zheng, Haonan and Bose, Sownak and Frenk, Carlos S. and Gao, Liang and Jenkins, Adrian and Liao, Shihong and Liu, Yizhou and Wang, Jie",
    title = "{The abundance of dark matter haloes down to Earth mass}",
    eprint = "2310.16093",
    archivePrefix = "arXiv",
    primaryClass = "astro-ph.GA",
    doi = "10.1093/mnras/stae289",
    journal = "Mon. Not. Roy. Astron. Soc.",
    volume = "528",
    number = "4",
    pages = "7300--7309",
    year = "2024"
}

@article{Paranjape:2012ks,
    author = "Paranjape, Aseem and Sheth, Ravi K.",
    title = "{Peaks theory and the excursion set approach}",
    eprint = "1206.3506",
    archivePrefix = "arXiv",
    primaryClass = "astro-ph.CO",
    doi = "10.1111/j.1365-2966.2012.21911.x",
    journal = "Mon. Not. Roy. Astron. Soc.",
    volume = "426",
    pages = "2789--2796",
    year = "2012"
}

@article{Paranjape:2012jt,
    author = "Paranjape, Aseem and Sheth, Ravi K. and Desjacques, Vincent",
    title = "{Excursion set peaks: a self-consistent model of dark halo abundances and clustering}",
    eprint = "1210.1483",
    archivePrefix = "arXiv",
    primaryClass = "astro-ph.CO",
    doi = "10.1093/mnras/stt267",
    journal = "Mon. Not. Roy. Astron. Soc.",
    volume = "431",
    pages = "1503--1512",
    year = "2013"
}

@article{Monaco:2001jg,
    author = "Monaco, P. and Theuns, T. and Taffoni, G.",
    title = "{Pinocchio: pinpointing orbit-crossing collapsed hierarchical objects in a linear density field}",
    eprint = "astro-ph/0109323",
    archivePrefix = "arXiv",
    doi = "10.1046/j.1365-8711.2002.05162.x",
    journal = "Mon. Not. Roy. Astron. Soc.",
    volume = "331",
    pages = "587",
    year = "2002"
}

@article{Taffoni:2001jh,
    author = "Taffoni, G. and Monaco, P. and Theuns, T.",
    title = "{Pinocchio and the hierarchical build-up of dark matter haloes}",
    eprint = "astro-ph/0109324",
    archivePrefix = "arXiv",
    doi = "10.1046/j.1365-8711.2002.05441.x",
    journal = "Mon. Not. Roy. Astron. Soc.",
    volume = "333",
    pages = "623",
    year = "2002"
}

@article{Dekker:2021scf,
    author = "Dekker, Ariane and Ando, Shin'ichiro and Correa, Camila A. and Ng, Kenny C. Y.",
    title = "{Warm dark matter constraints using Milky~Way satellite observations and subhalo evolution modeling}",
    eprint = "2111.13137",
    archivePrefix = "arXiv",
    primaryClass = "astro-ph.CO",
    doi = "10.1103/PhysRevD.106.123026",
    journal = "Phys. Rev. D",
    volume = "106",
    number = "12",
    pages = "123026",
    year = "2022"
}

@article{Ando:2024kpk,
    author = "Ando, Shin'ichiro and Horigome, Shunichi and Nadler, Ethan O. and Yang, Daneng and Yu, Hai-Bo",
    title = "{SASHIMI-SIDM: semi-analytical subhalo modelling for self-interacting dark matter at sub-galactic scales}",
    eprint = "2403.16633",
    archivePrefix = "arXiv",
    primaryClass = "astro-ph.CO",
    doi = "10.1088/1475-7516/2025/02/053",
    journal = "JCAP",
    volume = "02",
    pages = "053",
    year = "2025"
}

@article{Ondaro-Mallea:2023qat,
    author = {Ondaro-Mallea, Lurdes and Angulo, Raul E. and St{\"u}cker, Jens and Hahn, Oliver and White, Simon D. M.},
    title = "{Phase-space simulations of prompt cusps: simulating the formation of the first haloes without artificial fragmentation}",
    eprint = "2309.05707",
    archivePrefix = "arXiv",
    primaryClass = "astro-ph.GA",
    doi = "10.1093/mnras/stad3949",
    journal = "Mon. Not. Roy. Astron. Soc.",
    volume = "527",
    number = "4",
    pages = "10802--10821",
    year = "2023"
}

@article{Olea-Romacho:2025qag,
    author = "Olea-Romacho, Mar{\'\i}a Olalla and Fairbairn, Malcolm and Ralegankar, Pranjal",
    title = "{Can WIMPs Survive the Legacy of a Magnetised Early Universe?}",
    eprint = "2507.18692",
    archivePrefix = "arXiv",
    primaryClass = "hep-ph",
    month = "7",
    year = "2025"
}

@article{Ginat:2025kuz,
    author = "Ginat, Yonadav Barry and Nastac, Michael L. and Ewart, Robert J. and Konrad, Sara and Bartelmann, Matthias and Schekochihin, Alexander A.",
    title = "{Gravitational turbulence: The small-scale limit of the cold-dark-matter power spectrum}",
    eprint = "2501.01524",
    archivePrefix = "arXiv",
    primaryClass = "astro-ph.CO",
    doi = "10.1103/ychs-2d5p",
    journal = "Phys. Rev. D",
    volume = "112",
    number = "6",
    pages = "063501",
    year = "2025"
}

@article{Gao:2011rf,
    author = "Gao, L. and Frenk, C. S. and Jenkins, A. and Springel, V. and White, S. D. M.",
    title = "{Where will supersymmetric dark matter first be seen?}",
    eprint = "1107.1916",
    archivePrefix = "arXiv",
    primaryClass = "astro-ph.CO",
    doi = "10.1111/j.1365-2966.2011.19836.x",
    journal = "Mon. Not. Roy. Astron. Soc.",
    volume = "419",
    pages = "1721",
    year = "2012"
}

@article{Ando:2013ff,
    author = "Ando, Shin'ichiro and Komatsu, Eiichiro",
    title = "{Constraints on the annihilation cross section of dark matter particles from anisotropies in the diffuse gamma-ray background measured with Fermi-LAT}",
    eprint = "1301.5901",
    archivePrefix = "arXiv",
    primaryClass = "astro-ph.CO",
    doi = "10.1103/PhysRevD.87.123539",
    journal = "Phys. Rev. D",
    volume = "87",
    number = "12",
    pages = "123539",
    year = "2013"
}

@article{Polisensky:2015eya,
    author = "Polisensky, E. and Ricotti, M.",
    title = "{Fingerprints of the initial conditions on the density profiles of cold and warm dark matter haloes}",
    eprint = "1504.02126",
    archivePrefix = "arXiv",
    primaryClass = "astro-ph.GA",
    doi = "10.1093/mnras/stv736",
    journal = "Mon. Not. Roy. Astron. Soc.",
    volume = "450",
    number = "2",
    pages = "2172--2184",
    year = "2015"
}

@article{Ogiya:2017hbr,
    author = "Ogiya, Go and Hahn, Oliver",
    title = "{What sets the central structure of dark matter haloes?}",
    eprint = "1707.07693",
    archivePrefix = "arXiv",
    primaryClass = "astro-ph.CO",
    doi = "10.1093/mnras/stx2639",
    journal = "Mon. Not. Roy. Astron. Soc.",
    volume = "473",
    number = "4",
    pages = "4339--4359",
    year = "2018"
}

@article{White:2022yoc,
    author = "White, Simon D. M.",
    title = "{Prompt cusp formation from the gravitational collapse of peaks in the initial cosmological density field}",
    eprint = "2207.13565",
    archivePrefix = "arXiv",
    primaryClass = "astro-ph.CO",
    doi = "10.1093/mnrasl/slac107",
    journal = "Mon. Not. Roy. Astron. Soc.",
    volume = "517",
    number = "1",
    pages = "L46--L48",
    year = "2022"
}

@article{Ogiya:2016hyo,
    author = "Ogiya, Go and Nagai, Daisuke and Ishiyama, Tomoaki",
    title = "{Dynamical evolution of primordial dark matter haloes through mergers}",
    eprint = "1604.02866",
    archivePrefix = "arXiv",
    primaryClass = "astro-ph.CO",
    doi = "10.1093/mnras/stw1551",
    journal = "Mon. Not. Roy. Astron. Soc.",
    volume = "461",
    number = "3",
    pages = "3385--3396",
    year = "2016"
}

@article{Angulo:2016qof,
    author = "Angulo, Raul E. and Hahn, Oliver and Ludlow, Aaron and Bonoli, Silvia",
    title = "{Earth-mass haloes and the emergence of NFW density profiles}",
    eprint = "1604.03131",
    archivePrefix = "arXiv",
    primaryClass = "astro-ph.CO",
    doi = "10.1093/mnras/stx1658",
    journal = "Mon. Not. Roy. Astron. Soc.",
    volume = "471",
    number = "4",
    pages = "4687--4701",
    year = "2017"
}

@article{Wang:2025wki,
    author = "Wang, Yuchan and Bose, Sownak and Frenk, Carlos and Jenkins, Adrian",
    title = "{Buried but not destroyed: the evolution from prompt cusps to NFW haloes}",
    eprint = "2512.13804",
    archivePrefix = "arXiv",
    primaryClass = "astro-ph.CO",
    month = "12",
    year = "2025"
}

%% or
%% [B] Manual formatting (see below)
%% (i) We suggest to always provide author, title and journal data or doi:
%% in short all the informations that clearly identify a document.
%% (ii) please avoid comments such as "For a review'', "For some examples",
%% "and references therein" or move them in the text. In general, please leave only references in the bibliography and move all
%% accessory text in footnotes.
%% (iii) Also, please have only one work for each \bibitem.

\end{document}